\documentclass[aps,prl,twocolumn,floatfix,nofootinbib]{revtex4-1}
\usepackage{graphics}
\usepackage[normalem]{ulem} 
\usepackage{here}
\usepackage{bm}
\usepackage{subfigure}
\usepackage{graphicx}
\usepackage{amsthm}
\usepackage{notes2bib}
\usepackage{amsmath}
\usepackage{amssymb}
\usepackage{url}
\usepackage{enumerate}
\usepackage{epstopdf}
\usepackage{color}
\usepackage{mathtools}          
\usepackage{setspace}

\usepackage[usenames,dvipsnames,svgnames,table]{xcolor}
\usepackage{color}

\newcommand{\allblack}{\color{black}{}}

\newcommand{\mb}{\mathbf}

\usepackage{comment}
\begin{document}
\title{
{Data-driven modeling from biased small training data using periodic orbits}
}
\author{Kengo Nakai}
\affiliation{Graduate School of Environment, Life, Natural Science and Technology, 
   Okayama University   ,  Okayama 700-0082, Japan}
\author{Yoshitaka Saiki}
\affiliation{Graduate School of Business Administration, Hitotsubashi University, Tokyo 186-8601, Japan}

\begin{abstract}
In this study, we investigate the effect of reservoir computing training data on the reconstruction of chaotic dynamics. 
Our findings indicate that a training time series comprising a few periodic orbits of low periods can successfully reconstruct the Lorenz attractor.
We also demonstrate that biased training data does not negatively impact reconstruction success.
Our method's ability to reconstruct a physical measure is much better than the so-called cycle expansion approach, which relies on weighted averaging. 
Additionally, we demonstrate that fixed point attractors and chaotic transients can be 
accurately reconstructed by a model trained from a few periodic orbits, even when using different parameters.
\allblack
\end{abstract}
\date{\today}
\maketitle

\section{I. Introduction}
Reservoir computing, a brain-inspired machine-learning technique that employs a data-driven dynamical system, is effective in predicting time series and frequency spectra in chaotic behaviors, including fluid flow and global atmospheric dynamics~\cite{Verstraeten_2007,Zhixin_2017, Pathak_2017,Pathak_2018,Antonik_2018,nakai_2018,arcomano_2020,pandey_2020,huang_2020,kong_2021,kobayashi_2021,kobayashi_2024}.
\textcite{Pathak_2017} examined the Lorenz and Kuramoto--Sivashinsky systems and reported that the data-driven model obtained from reservoir computing could generate an arbitrarily long time series that mimics the dynamics of the original systems.\\ 
\indent To recognize the limitations of successful modeling using reservoir computing, we investigate the effect of selecting training data on modeling success. 
A constraint on training data is one of the major issues in data-driven modeling, and some studies have attempted to clarify related problems.
For example, Krishnagopal et al.~\cite{Krishnagopal_2018} investigated image classification problems on the MNIST handwritten digit dataset using reservoir computing. They found that prediction performance increased rapidly with training dataset size. 
It is also of great interest to see whether reservoir computing has the ability to predict a rare event that has not appeared in the past.\\ 
\indent In this study, we investigate the effect of biased small training data on the degree of dynamics reconstruction.
For this purpose, we use a set of periodic orbits as training data. 
A chaotic invariant set, including a chaotic attractor and a chaotic saddle, has a dense set of infinitely many periodic orbits, and a trajectory is considered to wander around periodic orbits.
We can easily create biased training datasets by selecting special periodic orbit types. 
We select a set of low-period periodic orbits localized in some regions or not passing through some regions.\\
\indent For a typical example, we employ the Lorenz system~\cite{l1963}:
\begin{equation}
\frac{dx}{dt}=10 (y-x),\  \frac{dy}{dt}=\rho x-y-xz,\ \frac{dz}{dt}=xy-\frac{8}{3}z.\label{eq:lorenz}
\end{equation}
A data-driven model is constructed from training data composed of a set of low-period periodic orbits of Eq.~(\ref{eq:lorenz}).
Two different parameter values $\rho=28$ and $23$ are considered; $\rho=28$ has a chaotic attractor, whereas $\rho=23$ has two fixed point attractors as well as a chaotic saddle, a nonattracting chaotic set~\cite{sparrow_1982}.
A set of periodic orbits of periods up to 9 is used as training data for reservoir computing.

\indent 
The remainder of this paper is organized as follows. 
The reservoir computing method is introduced in Section II. 
In Section III, employing the Lorenz system, we investigate the effect of reservoir computing training data composed of a set of low-period periodic orbits on the reconstruction of chaotic dynamics. 
Finally, Section VI presents the conclusions.

\section{II. Reservoir computing}
A reservoir is a recurrent neural network whose internal parameters are not adjusted to fit the data in the training process~\cite{Jaeger_2001,Jaeger_2004,Pathak_2017}. 
The reservoir can be trained by feeding it an input time series and fitting a linear/quadratic function of the reservoir state vector 
to the desired output time series. 
We do not use physical knowledge to construct a model. 
Our data-driven model using reservoir computing~\cite{ookubo_2024} is given by:
\begin{equation}
\begin{cases}
    \mb{u}(t)=\mb{W}^*_{\text{out}}\mb{r}(t)+ \mb{r}(t)^\text{T} \mb{W}^{*}_{\text{Qout}}\mb{r}(t) ,\\
	\mb{r}(t+\Delta t)=(1-\alpha)\mb{r}(t)+\alpha \tanh(\mb{A}\mb{r}(t)+\mb{W}_{\text{in}}\mb{u}(t)
	),
	\label{eq:reservoir}
	\end{cases}
	\end{equation}
where 
$\mb{u}(t) \in \mathbb{R}^M$ is a vector-valued variable, 
and its component is  
denoted as an output variable; 
$\mb{r}(t) \in \mathbb{R}^N~(N \gg M)$ is a reservoir state vector; 
$\mb{A} \in \mathbb{R}^{N\times N}$, 
$\mb{W}_{\text{in}} \in \mathbb{R}^{N\times M}$
and $\mb{W}^*_{\text{out}} \in \mathbb{R}^{M\times N}$
are matrices;
$\mb{W}^{*}_{\text{Qout}} \in \mathbb{R}^{M\times N\times N}$ is a tensor;
$\alpha$ ($0<\alpha\le 1$) is a coefficient; 
$\Delta t$ is a time step;
$\text{T}$ represents the transpose of a vector. 
We define $\tanh(\mb{v})=(\tanh(v_1), \tanh(v_2),\ldots,\tanh(v_N))^{\text{T}},$
for a vector $\mb{v} = (v_1,v_2,\ldots,v_N)^{\text{T}}$.

We now explain how to determine $\mb{W}^*_{\text{out}}$ and $\mb{W}^*_{\text{Qout}}$ in (\ref{eq:reservoir}). 
The time development of the reservoir state vector 
$\mb{r}(l \Delta t)$ is determined by the second equation of (\ref{eq:reservoir}), with training time series data $\{\mb{u}(l \Delta t)\} (-L_0\le l \le L)$, where $L_0$ is the transient time and  $L$ is the time length to determine $\mb{W}^*_{\text{out}}$ and $\mb{W}^*_{\text{Qout}}$. 
For given random matrices 
$\mb{A}$ and $\mb{W}_{\text{in}}$, 
we determine $\mb{W}_{\text{out}} \in \mathbb{R}^{M\times N}$ and $\mb{W}_{\text{Qout}} \in \mathbb{R}^{M\times N\times N}$ so that the following relation holds: 
\begin{align}
    \mb{W}_\text{out}\mb{r}(l\Delta t)
    + \mb{r}(l\Delta t)^{\text{T}}\mb{W}_{\text{Qout}}\mb{r}(l\Delta t)
    \approx \mb{u}(l\Delta t).
    \nonumber 
\end{align}
The detail for training time series along periodic orbits is described in the Supplemental Material. 
Remark that the similar results can be obtained by modeling with a linear form instead of a quadratic form.

\begin{table}[htb]
\allblack
\small
	\begin{tabular}{|l|r|r|}
	\hline
    Dimension $M$ of input and output variables& 6 \\  \hline
	Dimension $N$ of reservoir state vector & 50 \\  \hline
    Time step $\Delta t$ for a model~(\ref{eq:reservoir}) & 0.01 \\  \hline
    Maximal eigenvalue of $\mb{A}$ & 1 \\  \hline
    Nonlinearity degree $\alpha$ in a model~(\ref{eq:reservoir})&0.3  \\ \hline
    Delay time $\Delta \tau$ for input and output variables &0.05 \\ \hline
    \end{tabular}
    \caption{{\bf The list of parameters and their values used in the reservoir computing.}
    We use $\mb{u}(t) = (x(t),y(t),z(t),x(t-\Delta \tau),y(t-\Delta \tau),z(t-\Delta \tau))$ for the input variable, where $\Delta\tau$ is the delay time.
 	}
 	\label{tab:parameter}
\end{table}
\section{III. Results} 
\indent We construct a data-driven model for the Lorenz system~\eqref{eq:lorenz} with the parameter $\rho=28$, unless otherwise denoted.  
We continue to use the same 
matrices $\mb{A}$ and $\mb{W}_{\text{in}}$ for all cases, 
meaning that we only train $\mb{W}_{\text{out}}$ and $\mb{W}_{\text{Qout}}$ using each 
set of unstable periodic orbits, including the case of $\rho = 23$.
See Table~\ref{tab:parameter} for the set of parameters.
%
\subsection{Modeling using only three periodic orbits}
\subsubsection{$\rho=28$: the case for the chaotic attractor} 
First, we investigate the case of $\rho=28$, where the system has a chaotic attractor. 
We construct a data-driven model by training the three periodic orbits with the low periods for the Lorenz system with $\rho=28$ (See Fig.~\ref{fig:three-periodic-orbit}(a)). 
The periodic orbits consist of one period-2 and two period-3 orbits, where the period denotes the integer period of the corresponding Poincar\'e map. 
The two period-3 orbits are symmetric with each other under the transformation $(x,y,z)\mapsto(-x,-y,z).$
Consequently, we predict chaotic time series data of the Lorenz system using the constructed model. 
Figure~\ref{fig:three-periodic-orbit} (b) shows a three-dimensional plot of a model trajectory, time evolution of the $x$ variable, and return plots $(z_n, z_{n+1})$, where $z_n$ denotes the $n$-th local maximum of the $z$ variable. 

We confirm the reproduction of chaotic behavior 
of the dynamical system using a data-driven model, which 
is obtained by training the periodic orbits of the system. 
\subsubsection{$\rho=23$: the case for the fixed point attractor} 
\begin{figure*}[ht]
    \begin{minipage}{0.35\hsize}
        (a) training data ($\rho=28$)\\
        \vspace{-5mm}
            \includegraphics[width=0.49\columnwidth,height=35mm]{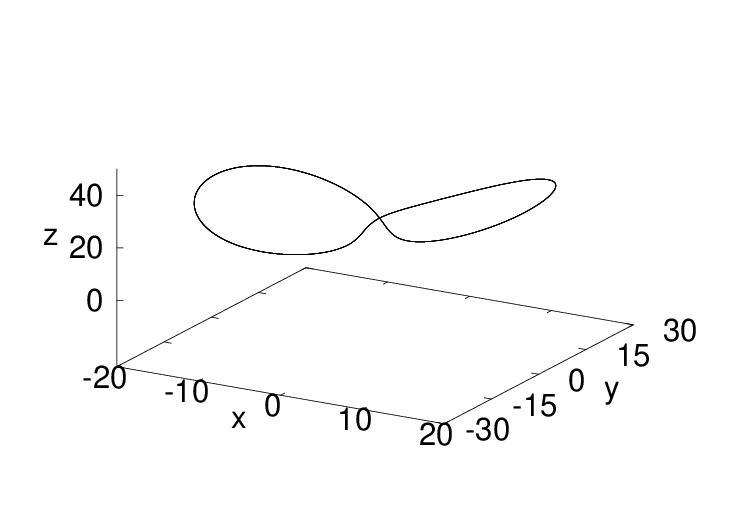}\\
            \vspace{-14mm}
            \includegraphics[width=0.49\columnwidth,height=35mm]{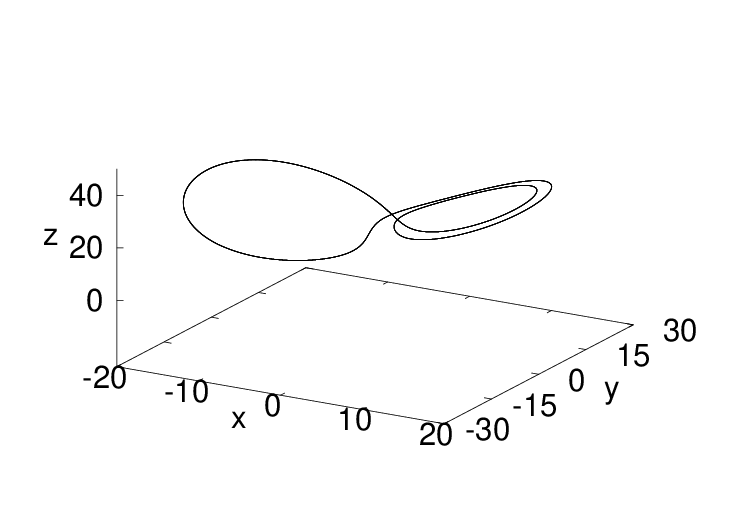}
            \includegraphics[width=0.49\columnwidth,height=35mm]{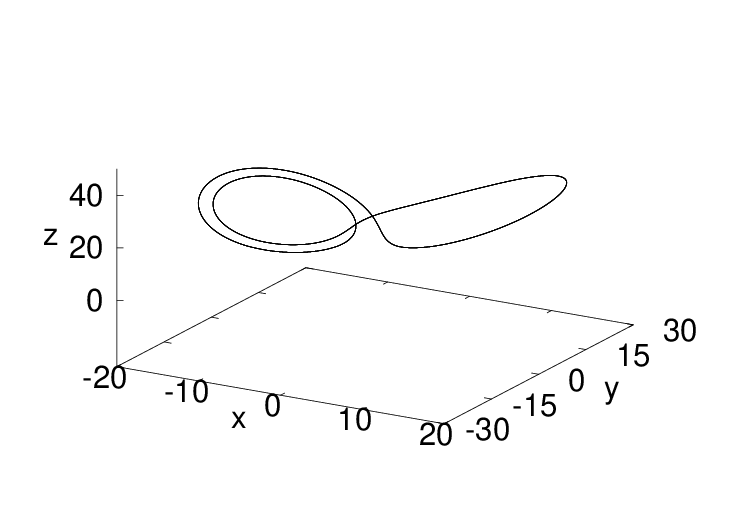}
    \end{minipage}
    \begin{minipage}{0.08\hsize}
		{\Huge $\Rightarrow$ }
    \end{minipage}
    \begin{minipage}{0.55\hsize}
        (b) reservoir model\\
        \begin{minipage}{0.4\hsize}
            \begin{center}
	   		      \includegraphics[width=0.99\columnwidth,height=35mm]{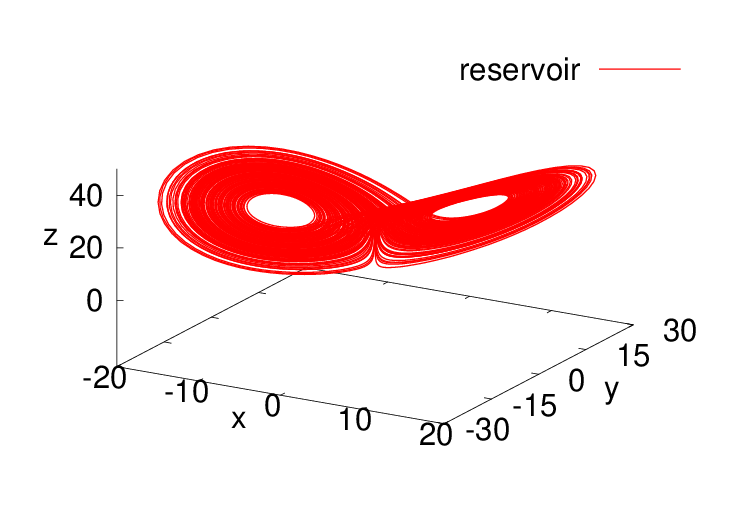}\\
                \vspace{-2mm}
		          \includegraphics[width=0.99\columnwidth,height=30mm]{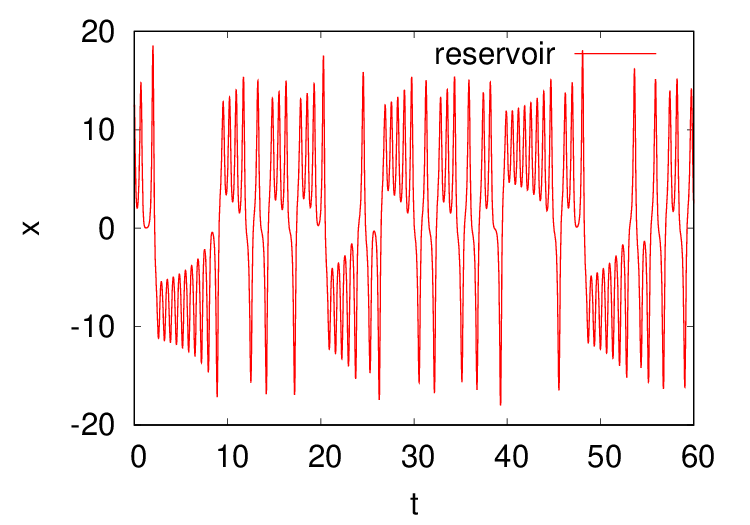}
		      \end{center}
        \end{minipage}
                \hspace{-15mm}
        \begin{minipage}{0.57\hsize}
            \begin{center}
                \vspace{6mm}
                \includegraphics[width=1.3\columnwidth]{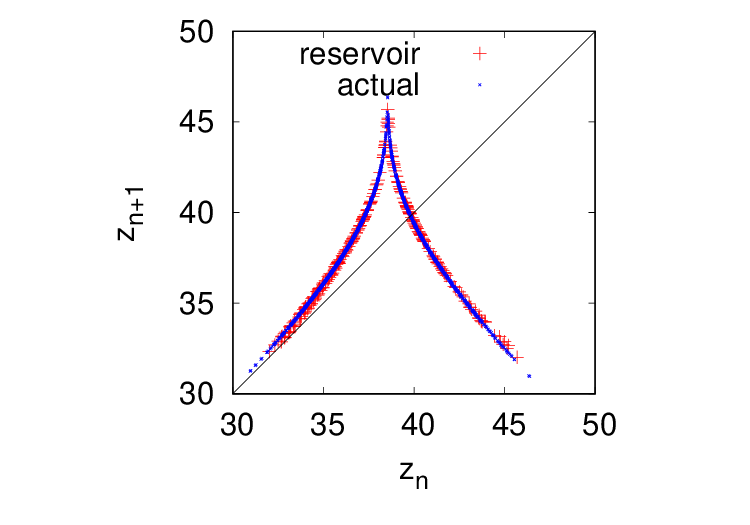}
		      \end{center}
        \end{minipage}
	\end{minipage}
\\
    \vspace{7mm}
    \begin{minipage}{0.35\hsize}
        (c) training data ($\rho=23$)\\
        \vspace{-5mm}
            \includegraphics[width=0.49\columnwidth,height=35mm]{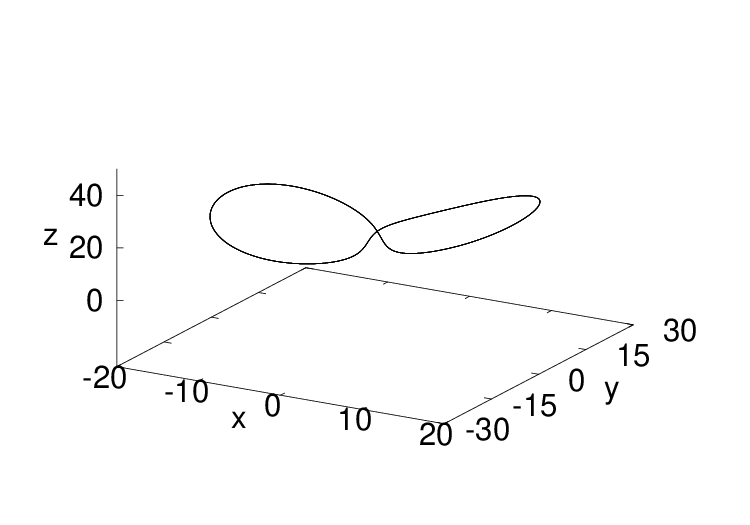}\\
            \vspace{-14mm}
            \includegraphics[width=0.49\columnwidth,height=35mm]{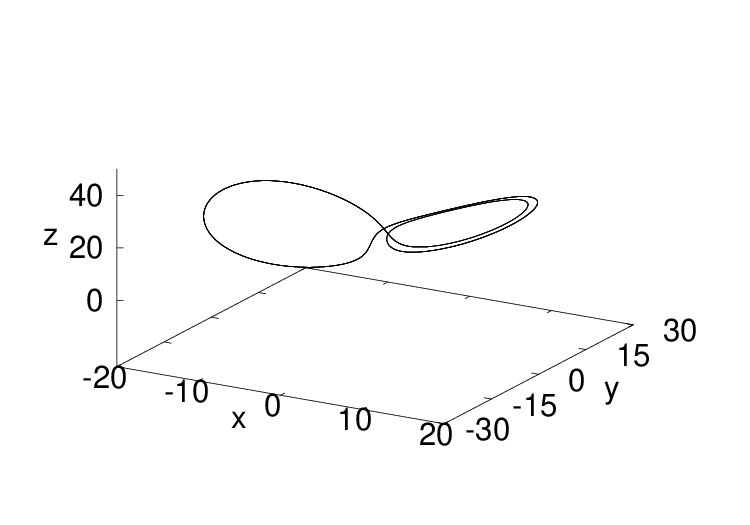}
            \includegraphics[width=0.49\columnwidth,height=35mm]{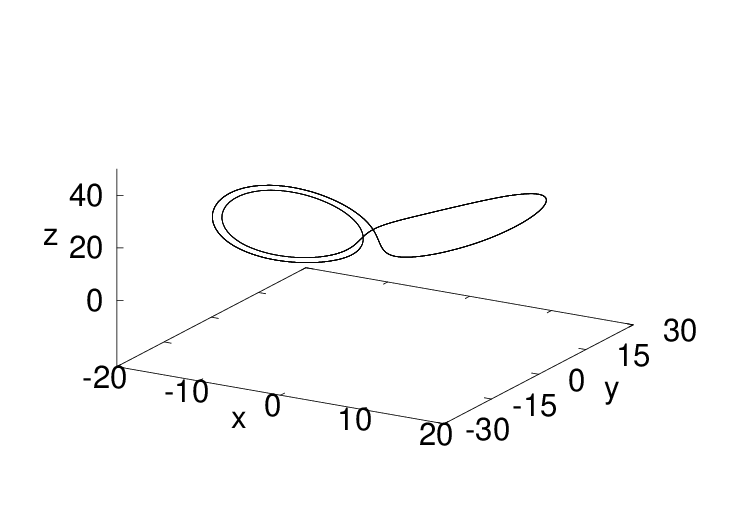}
    \end{minipage}
    \begin{minipage}{0.08\hsize}
		{\Huge $\Rightarrow$ }
    \end{minipage}
    \begin{minipage}{0.55\hsize}
        (d) reservoir model\\
        \begin{minipage}{0.4\hsize}
            \begin{center}
	   		      \includegraphics[width=0.99\columnwidth,height=35mm]{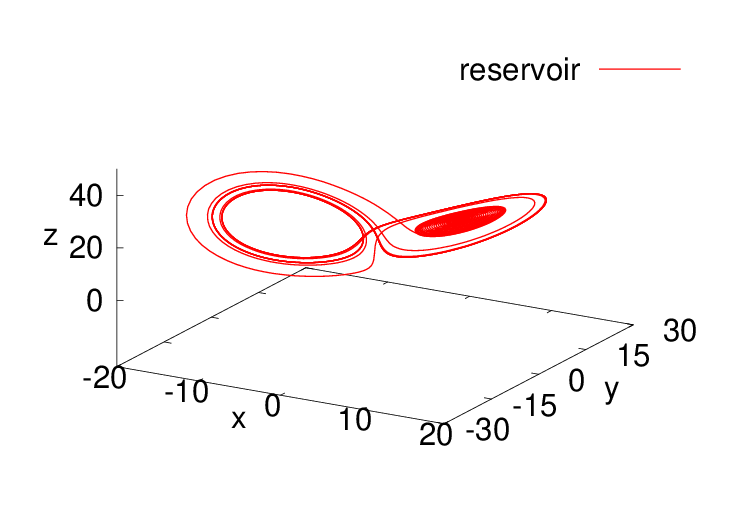}\\
                \vspace{-2mm}
		          \includegraphics[width=0.99\columnwidth,height=30mm]{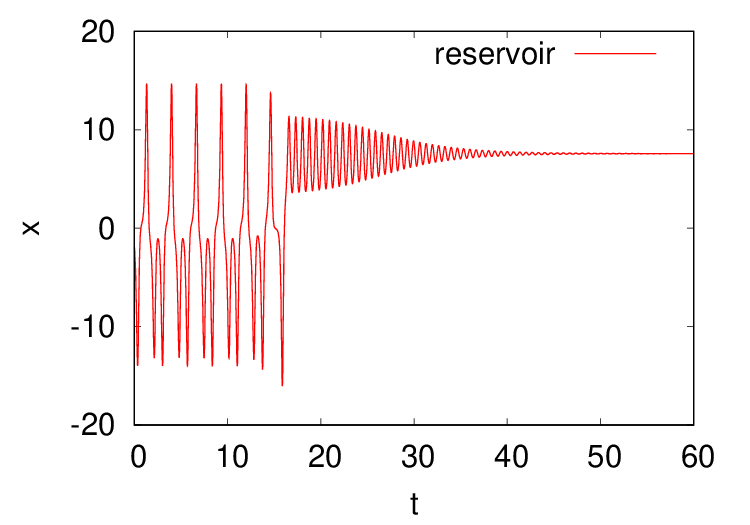}
		      \end{center}
        \end{minipage}
                \hspace{-15mm}
        \begin{minipage}{0.57\hsize}
            \begin{center}
                \vspace{6mm}
                \includegraphics[width=1.3\columnwidth]{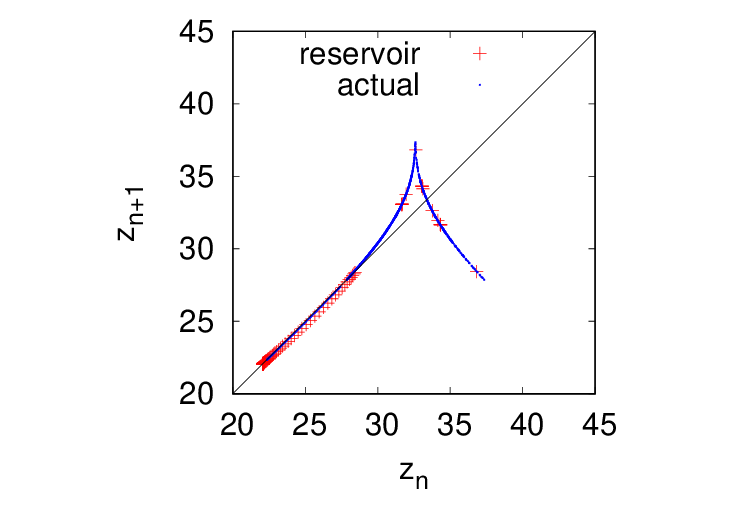}
		      \end{center}
        \end{minipage}
	\end{minipage}
\caption{{\bf Three periodic orbits model the Lorenz dynamics, chaotic attractor and chaotic saddle with two sets of parameters.}
We construct data-driven models by training the three periodic orbits of period-2 and period-3 for $\rho=28$ (top) and $\rho=23$ (bottom). Each reservoir model reconstructs the dynamics; for $\rho=28$, the model has the chaotic attractor, and for $\rho=23$, the model has a chaotic transient as well as the fixed point attractors. 
Figure~\ref{fig:convergence} (b) shows the density distribution of the model for $\rho=28$. 
The model orbit for $\rho=23$ converges to a point $(7.575, 7.518, 22.069)$. 
Depending on the initial condition, a trajectory can also be attracted to another fixed point $(-7.636,-7.613,21.978)$. 
The coordinates of the two actual attracting fixed points are approximately $(\pm 7.659, \pm 7.659, 22.000)$. 
}
\label{fig:three-periodic-orbit}
\end{figure*}

Next, we construct a data-driven model by training three periodic orbits with low periods with $\rho=23$ (Fig.~\ref{fig:three-periodic-orbit} (c)). 
Figure~\ref{fig:three-periodic-orbit} (d) shows a three-dimensional plot of a model trajectory,
time evolution of the $x$ variable, and return plots $(z_n, z_{n+1})$. 
We confirm the reproduction of the dynamical system using a data-driven model, which is obtained by training the system's periodic orbits. 
Despite using the same matrices $\mb{A}$ and $\mb{W}_{\text{in}}$, the data-driven models capture  differences between 
$\rho=28$ and $\rho=23$; 
the former has a chaotic attractor, and the latter has two fixed-point attractors. Before approaching the fixed points, dynamics shows a chaotic transient wandering near the chaotic saddle.

\subsection{Convergence to the actual density distribution}
\begin{figure}
\centering
\begin{minipage}{1\linewidth}
\centering
		\subfigure[]{%
		\includegraphics[width=0.48\columnwidth,height=0.45\columnwidth]{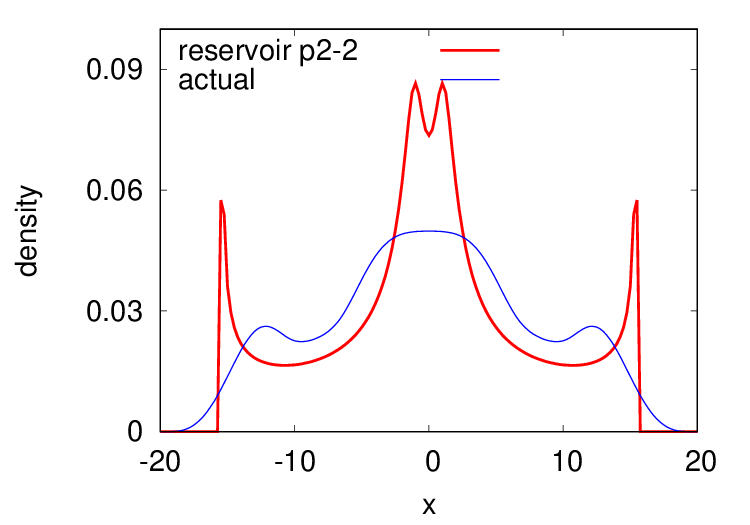}
		}
		\subfigure[]{%
		\includegraphics[width=0.48\columnwidth,height=0.45\columnwidth]{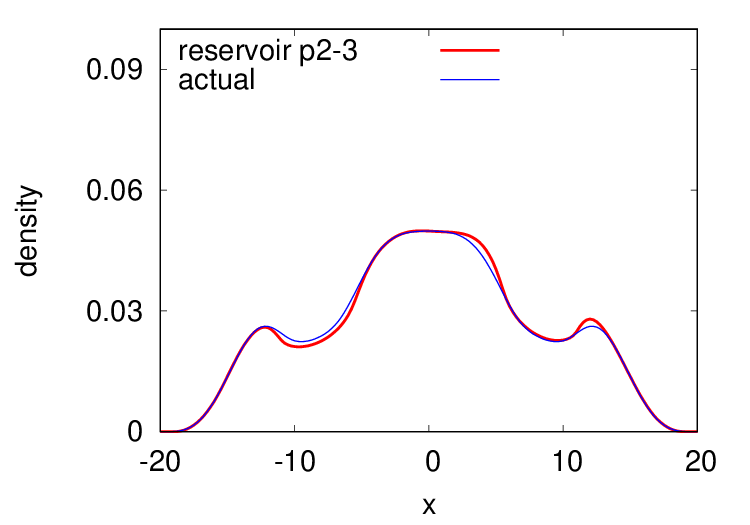}
		}\\%
		\subfigure[]{%
		\includegraphics[width=0.48\columnwidth,height=0.45\columnwidth]{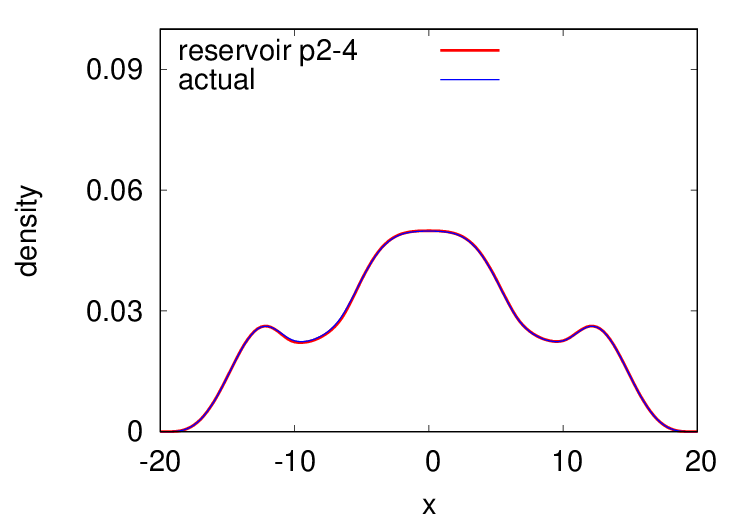}
		}%
		\subfigure[]{%
		\includegraphics[width=0.48\columnwidth,height=0.45\columnwidth]{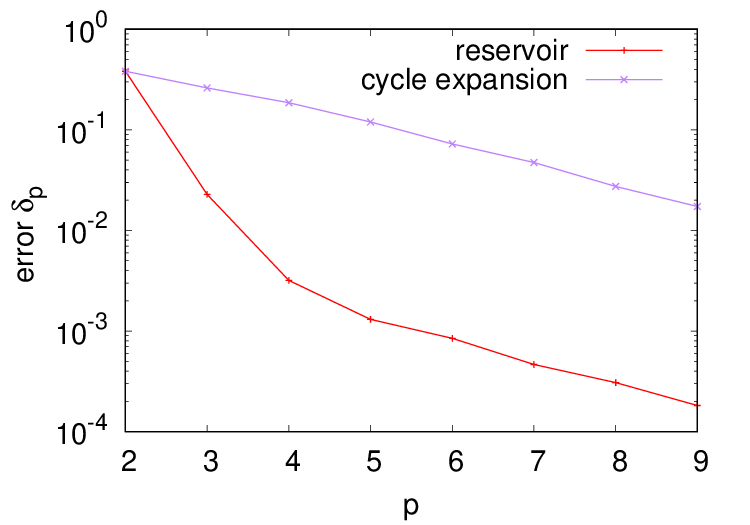}
		}%
    \end{minipage}
\caption{{\bf Fast convergence of the density distribution $\Phi_p$ to the actual one $\Phi_\text{Lorenz}$.} A density distribution is created using each of the three models trained from different sets of periodic orbits. The model in panel (a) is trained from a periodic orbit of period-2,
(b) periodic orbits of period-2 and 3, and (c) period-2, 3 and 4.
For (a), the model has a period-2 periodic attractor; this is because derivative information cannot be trained from the period-2 training orbit. 
In panel (d), the error $\delta_p$ of the density distribution compared to the actual one is shown with respect to $p$, where 
$p$ is the maximum value of the period of periodic orbits used for the modeling. The error $\delta_p$ is compared with the so-called ``cycle expansion," a weighted averaging technique for approximating the actual density distribution $\Phi_\text{Lorenz}$. Note that the number of periodic orbits grows exponentially as the period $p$ increases. 
}
\label{fig:convergence}
\end{figure}

We construct data-driven models by training periodic orbits of period-$2$ to period-$p$ ($p=2,3,\ldots,9$).
For each constructed model, we created a density distribution, $\Phi_p$, of the $x$ variable. 
Figure~\ref{fig:convergence} (a-c) shows $\Phi_{p}$ for $p=2,3,4$ along with the density distribution of the actual Lorenz system, 
denoted by 
$\Phi_{\text{Lorenz}}$.
The density distribution $\Phi_{2}$ significantly
differs from $\Phi_{\text{Lorenz}}$ but $\Phi_{3}$ approximates $\Phi_{\text{Lorenz}}$, suggesting that only three periodic orbits of low periods are practically 
sufficient to reconstruct the Lorenz attractor. 
We compute the error between 
the density distribution $\Phi_{p}$ and
the actual distribution $\Phi_{\text{Lorenz}}$
as follows:
$$\delta_p=\int_{-\infty}^{\infty} |\Phi_{p}(x)-\Phi_{\text{Lorenz}}(x)| dx.$$
The error $\delta_p$ decreases as $p$ increases (Fig.~\ref{fig:convergence}(d)). 
To evaluate the convergence speed, we compare it with  the so-called ``cycle expansion,'' a weighted averaging technique 
along periodic orbits~\cite{grebogi88,cvitanovic93,christiansen97,zoldi98,dhamala_1999,lai_1997}. 
Figure~\ref{fig:convergence}(d) shows that the convergence speed using reservoir computing is much faster. 
These results suggest that reservoir computing interpolates the dynamical system properties more effectively than cycle expansion employing periodic orbits with their stability exponents and periods.

\subsection{Case 1: Reproducibility when removing some periodic orbits from training data} 
\begin{figure}
    \begin{minipage}{0.47\hsize}
        \begin{center}
        ~~training data \\
        \includegraphics[width=0.99\columnwidth]{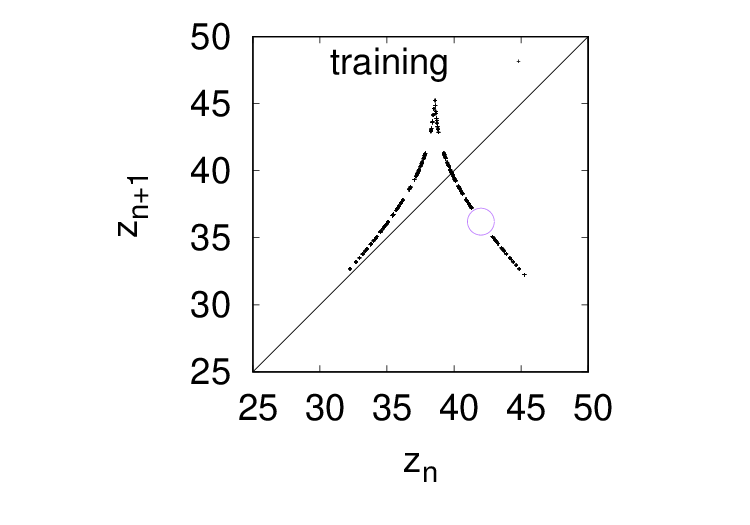}
        \includegraphics[width=0.99\columnwidth]{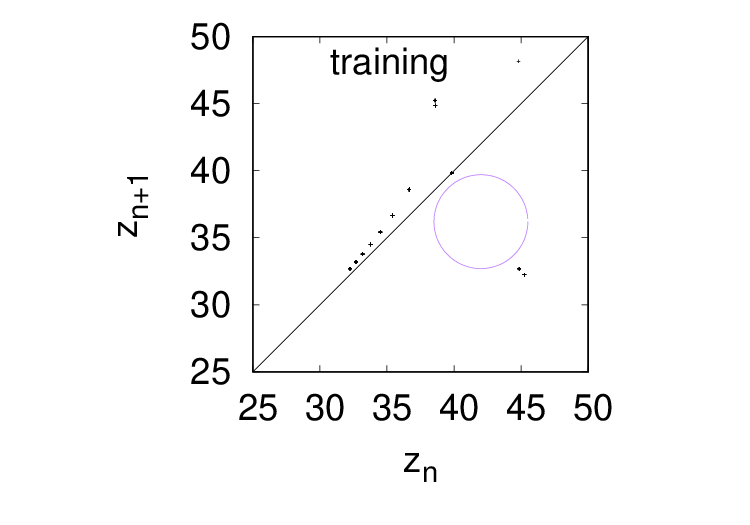} 
        \end{center}
    \end{minipage}
    \begin{minipage}{0.03\hsize}
		{\LARGE $\Rightarrow$ }\\\vspace{20mm}
		{\LARGE $\Rightarrow$ }
    \end{minipage}
    \begin{minipage}{0.47\hsize}
        \begin{center}
        ~~reservoir model \\
        \includegraphics[width=0.99\columnwidth]{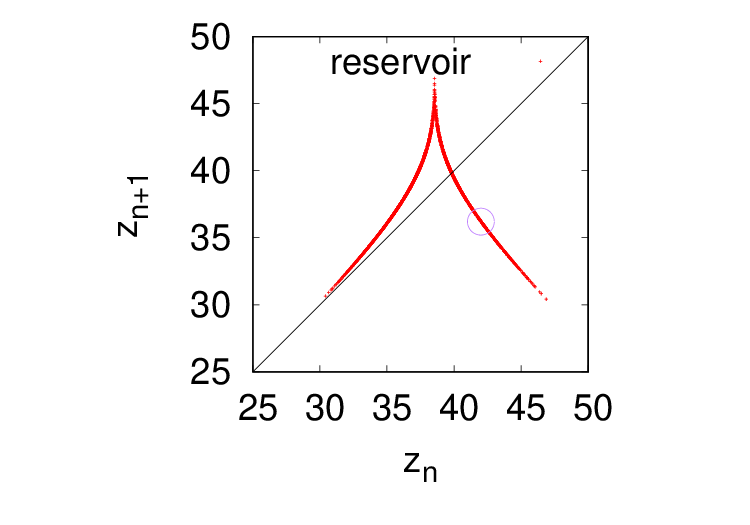}
        \includegraphics[width=0.99\columnwidth]{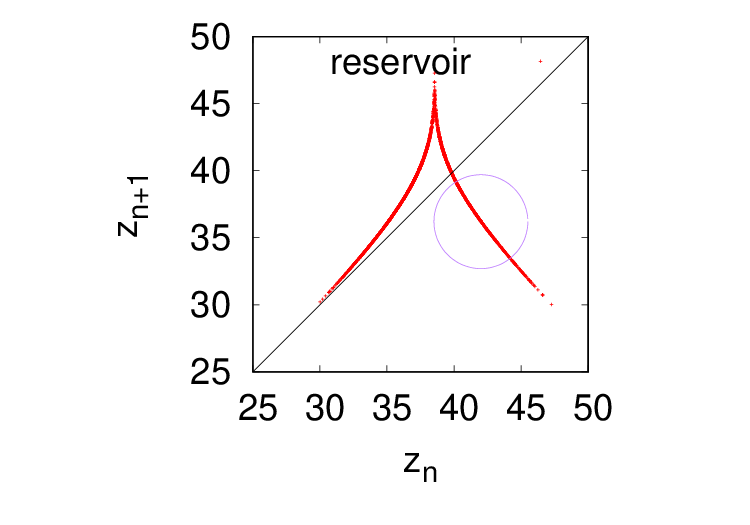}     
        \end{center}
	\end{minipage}
\caption{{\bf Modeling from biased training data (Case 1).}
Each left panel shows training data in the return plots defined by the maximal value of the $z$ variable, and the right panel shows the corresponding return plots created from the reservoir model.
Even if some periodic orbits that pass through a certain region denoted by the circle are excluded from the training dataset, the reservoir model reconstructs the attractor.
The regions for excluding periodic orbits are denoted by the circles: $B_{1}(42,36.2)$ (top) and  $B_{3.5}(42,36.2)$ (bottom). 
}
\label{fig:return-map}
\end{figure}

We investigate the effect of biased training data on modeling. In Case 1, we create a training dataset by removing certain periodic orbits from the list of periodic orbits with periods up to 9. 
If a periodic orbit passes through a ball $B_{d}(42,36.2)$ with radius $d=1$ centered at $(z_n,z_{n+1})=(42, 36.2)$, we remove the orbit from the list of periodic orbits. 
Figure~\ref{fig:return-map} shows the return plots of the training periodic orbits
in the list, and 
those generated by the data-driven model.
Our results show that the model successfully reconstructs the information within the ball $B_{1}(42,36.2)$.
A similar result can be obtained when we select a different center point and/or radius of a ball.
The case for a ball $B_{3.5}(42,36.2)$ is also shown in Fig.~\ref{fig:return-map}. 

\subsection{Case 2: Reproducibility when some periodic orbits are repeatedly trained multiple times} 
\begin{figure}[H]
		\subfigure[]{%
		\includegraphics[width=0.43\columnwidth,height=0.43\columnwidth]{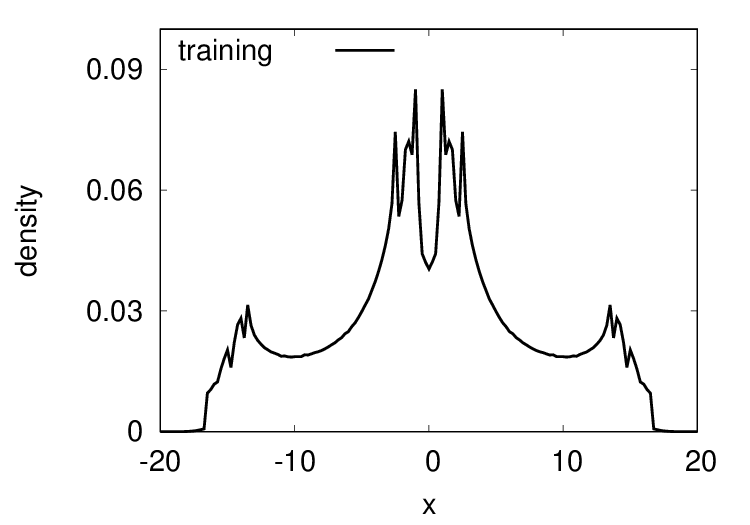}
		}%
		\subfigure[]{%
		\includegraphics[width=0.43\columnwidth,height=0.43\columnwidth]{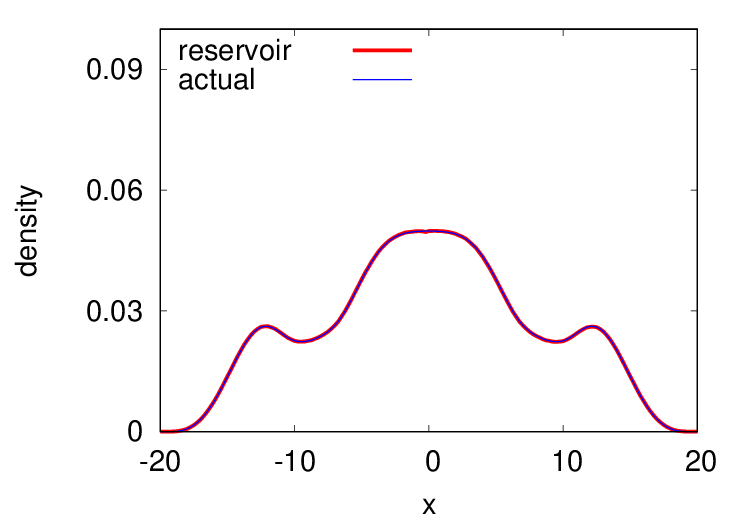}
		}%

\caption{{\bf Modeling from biased training data (Case 2)} 
The left panel (a) shows the density distribution of the training data, and the right (b) shows the density distribution of the reservoir model.
Even if some periodic orbits that pass through a certain region denoted by the circle are trained multiple times, the reservoir model generates a density distribution $\Phi_\text{bias}$ which agrees with $\Phi_\text{Lorenz}$.
Each of the periodic orbits that pass through $B_{0.05}(42,36.2)$ is trained additionally 100 times. 
}
\label{fig:additional}
\end{figure}

We investigate the effect on the data-driven model by repeatedly training selected periodic orbits in Case 2.
We select periodic orbits of period-2 to period-9 
passing through a ball $B_{d}(42,36.2)$ on the $(z_n,z_{n+1})$ plane, similar to Case 1. 
When $d=0.05$, we select three periodic orbits of period-7, 8 and 9.
After training periodic orbits of period-2 to period-9, we train each of the selected three periodic orbits $100$ times. 
Figure~\ref{fig:additional}(a) shows the density distribution of the $x$ variable of the training data, which has a skewed shape due to the biased training data.
Figure~\ref{fig:additional}(b) shows the density distribution $\Phi_{\text{bias}}$ of the data-driven model and that of the actual Lorenz system $\Phi_{\text{Lorenz}}$. 
Our results demonstrate that the density distribution of the data-driven model does not change even when certain periodic orbits are trained multiple times.

\section{VI. Concluding remarks}
In this study, we investigated the reconstruction of chaotic dynamics using small biased training data composed of periodic orbits. 
We found that a time series comprising a small number ($\approx 10^0\sim 10^2$) of low-period periodic orbits is sufficient to construct a reservoir model reconstructing actual chaotic dynamics.   
Our results show that the difference in density distributions between the constructed data-driven models and the actual Lorenz system decreased much faster than that obtained using the so-called cycle expansion approach.
Even when certain periodic orbits passing through some regions were removed from the training data, the data-driven model could interpolate this missing information.  The obtained model accurately reconstructs the actual statistical properties.
The same result holds 
even when certain periodic orbits were
additionally trained multiple times.
Based on our investigation, we conjectured that some rare events can be 
reconstructed from a time series of typical time periods using reservoir computing, even when such rare events are not observed in the training time series. 

\section*{Acknowledgements}
KN was supported by JSPS KAKENHI Grant No.22K17965. 
YS was supported by JSPS KAKENHI Grant No.19KK0067, 21K18584, 23H04465 and 24K00537.
The computation was performed using the JHPCN (jh240051) and the Collaborative Research Program for Young $\cdot$ Women Scientists of ACCMS and IIMC, Kyoto University. 
\bibliographystyle{apsrev4-2}

\end{document}